\documentclass[]{spie}  
\usepackage[]{graphicx}

\title{Near-infrared wavefront sensing for the VLT interferometer } 

\author{Stefan Hippler\supit{a}, Wolfgang Brandner\supit{a}, Yann Cl\'enet\supit{b}, Felix Hormuth\supit{a}, Eric Gendron\supit{b}, \\ Thomas Henning\supit{a},
Ralf Klein\supit{a}, Rainer Lenzen\supit{a}, Daniel Meschke\supit{a}, Vianak Naranjo\supit{a}, \\ Udo Neumann\supit{a}, 
Jos\'{e} Ricardo Ramos\supit{a}, Ralf-Rainer Rohloff\supit{a}, Frank Eisenhauer\supit{c}
\skiplinehalf
\supit{a}Max-Planck-Institut f\"ur Astronomie, K\"onigstuhl 17, D-69117 Heidelberg, Germany; \\
\supit{b}Observatoire de Paris \`a Meudon,  5, place Jules Janssen, 92195 Meudon, France;\\
\supit{c}Max-Planck-Institut f\"ur Extraterrestrische Physik, Giessenbachstra{\ss}e, D-85748 Garching, Germany
}

\authorinfo{Further author information: (Send correspondence to Stefan Hippler), E-mail: hippler@mpia.de}
 
\begin{document} 
  \maketitle 

\begin{abstract}
The very large telescope (VLT) interferometer (VLTI) in its current operating state is equipped with high-order 
adaptive optics (MACAO) working in the visible spectrum. A low-order near-infrared wavefront sensor (IRIS) is available to measure 
non-common path tilt aberrations downstream the high-order deformable mirror. For the next generation 
of VLTI instrumentation, in particular for the designated GRAVITY instrument, we have examined various designs 
of a four channel high-order near-infrared wavefront sensor. Particular objectives of our study were the 
specification of the near-infrared detector in combination with a standard wavefront sensing system. In this paper we 
present the preliminary design of a Shack-Hartmann wavefront sensor operating in the near-infrared wavelength 
range, which is capable of measuring the wavefronts of four telescopes simultaneously. 
We further present results of our design study, which  aimed at providing a first instrumental concept for GRAVITY.
\end{abstract}



\section{INTRODUCTION}
\label{sec:intro}  
GRAVITY\cite{Gillessen2006a} is an interferometric imager with 10\,$\mu$as astrometric capability, which will coherently combine the light from all four ESO VLT 8m unit telescopes. 
GRAVITY will carry out the ultimate empirical test whether or not the Galactic Centre harbours a 4 million solar mass black hole, without resorting to theoretical assumptions. If the current interpretation of the near-infrared (NIR) flares from SgrA* is correct, GRAVITY has the potential of directly determining the space-time metric around this black hole. The instrument will also be able to unambiguously detect and measure the mass of black holes in massive star clusters throughout the Milky Way and in many active galactic nuclei to z$\sim$0.1. It will make unique measurements on gas jets in young stellar objects and active galactic nuclei. It will explore binary stars, exo-planet systems and young stellar disks. Because of its superb sensitivity GRAVITY will excel in milli-arcsecond phase-referenced imaging of faint celestial sources of any kind. Because of its outstanding astrometric capabilities, it will detect motions throughout the local Universe and perhaps beyond. Because of its spectroscopic and polarimetric capabilities it is capable of detecting gas motions and magnetic field structures on sub-milliarcsecond scales. 

GRAVITY will measure fringe visibility magnitudes and fringe positions with high precision in order to achieve its goals.
The visibility magnitude signal-to-noise ratio depends, among others, on the optical quality of the wavefronts brought to interference. The seeing-limited optical quality can be enhanced with the help of adaptive optics. Adaptive optics requires sufficiently bright reference stars to work properly. The VLTI MACAO\cite{MACAO2004} systems provide adaptive optics using the visible light of reference stars. For bright, yet highly embedded (reddened) targets, as found, e.g., in the galactic center region, wavefront sensing at near-infrared wavelengths is required as long as laser guide stars are not available. Thus GRAVITY will employ near-infrared wavefront sensors, which in combination with the existing MACAO deformable mirrors form a complete near-infrared sensitive adaptive optics system for the VLTI\cite{Schoeller2007} and its suite of instruments.

The selected baseline concept for the infrared wavefront sensor is a Shack Hartmann system. It will be located in the VLTI laboratory, thus also sensing tunnel seeing (see section~\ref{sec:tunnel}) in the VLTI optical train. The wavefront correction will be applied to the MACAO deformable mirrors located at the Coud\'e focal station of each unit telescope. The main requirements for this near-infrared adaptive optics system are the delivery of a 36\% Strehl point spread function in K-band while guiding on a K=7 star. We discuss the pros and cons of using a single detector for all four telescopes rather than one detector per telescope. The requirements for the near-infrared detector are discussed and compared with available devices. Since pupil stability is an important issue for the VLTI, we present a scheme to measure possible pupil drifts in real-time with the wavefront sensor.
   \begin{figure}[ht]
   \begin{center}
   \begin{tabular}{c}
   \includegraphics[width=16cm]{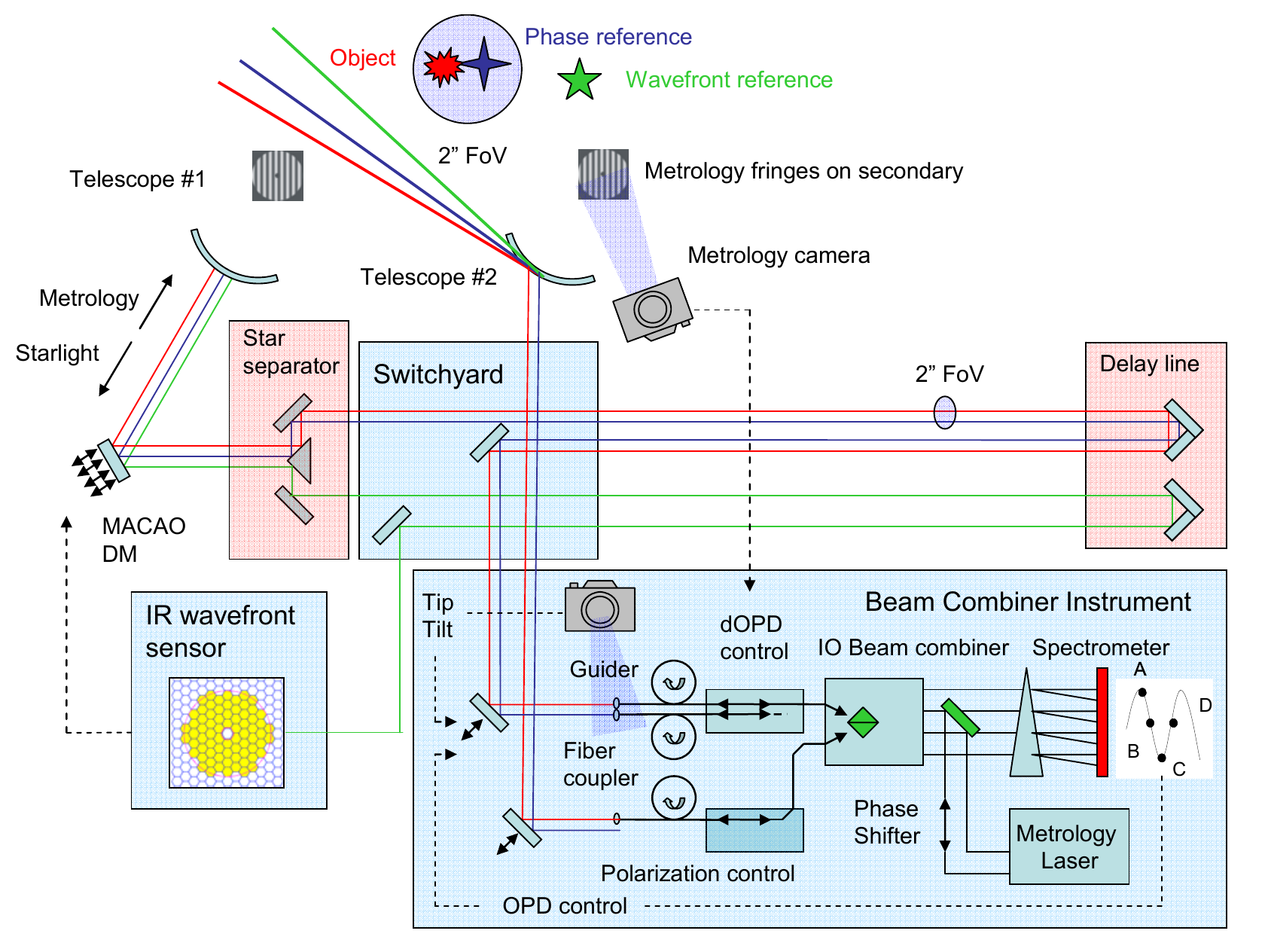}
   \end{tabular}
   \end{center}
   \caption[Schematic] 
   { \label{fig:Schematic} 
Schematic overview of GRAVITY.}
   \end{figure} 
\section{Adaptive optics requirements for GRAVITY}
The requirements on the adaptive optics (AO) system resulting from the GRAVITY science cases can be summarized as follows:
\begin{itemize}
\item provide wavefront correction
\item provide NIR wavefront sensing
\item allow for off-axis wavefront sensing
\item wavefront sensor wavelength range: H- and K-band
\end{itemize}

Image quality in terms of Strehl ratio (SR) requirements as a function of reference star brightness in K-band assuming median seeing conditions on Paranal are:
\begin{itemize}
\item m$_K$ = 7: SR=0.25 at 45 degrees zenith distance and 7 arcsec distance to the reference star 
\item m$_K $= 7: SR= 0.4  at 45 degrees zenith distance on-axis 
\item m$_K$ = 10: SR=0.1 at zenith and on-axis 
\end{itemize}

\begin{figure}[ht]
   \begin{center}
   \begin{tabular}{c}
   \includegraphics[width=14cm]{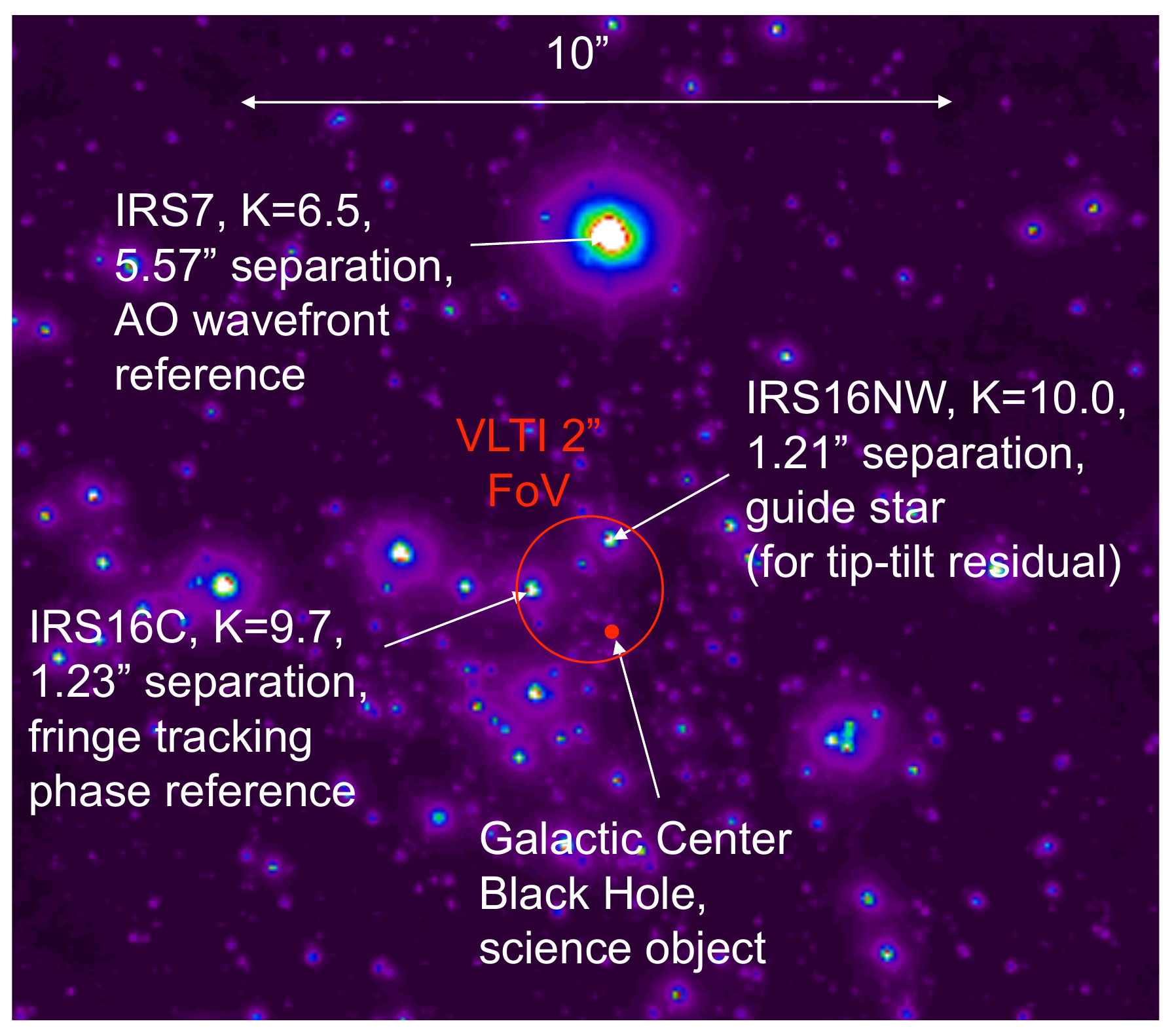}
   \end{tabular}
   \end{center}
   \caption[Schematic] 
   { \label{fig:GC_AO} 
The GRAVITY galactic center observing field with guide stars.}
   \end{figure}

A possible galactic center scenario for observations with GRAVITY is shown in fig.~\ref{fig:GC_AO}.

\section{System baseline}
\subsection{Wavefront sensor options}
In our design study for GRAVITY we have selected as baseline system a Shack-Hartmann type wavefront sensors (SHWFS) because of their robustness and uncomplicated assembly. In addition, and unlike, e.g., curvature or pyramid sensors, they provide a direct image of the reference source, which provides a useful diagnostics both during the commissioning as well as the science operation of the NIR AO system. Pyramid WFS are in general optimized to focus the light from a single star on the tip of the pyramid. Thus they are probably less optimal for wavefront sensing in crowded regions, like, e.g., the centers of starburst clusters, or in the case of observations of (close) binary stars.

\subsection{Detector options}
The selection of the baseline for the NIR wavefront sensor (WFS) detector for GRAVITY was mostly based on availability, and of course technical requirements.  The baseline design follows the line of injecting all 4 telescope beams into 1 single detector rather than into 4 separate ones.   Table~\ref{tab:detreq} shows the requirements of the IR WFS Detector on which the selection process was based. 

\begin{table}[ht]
\begin{center}
\begin{tabular}{|p{1.5in}|p{2.0in}|p{2.0in}|} \hline 
\textbf{Parameter} & \textbf{Desired Characteristics} & \textbf{Minimum Requirements} \\ \hline 
Operating Range & 1 -- 2.5 µm & 1.45 -- 2.45 µm \\ \hline 
Frame size & 4 x 100 x 100 & 4 x 60 x 6 x 6 \\ \hline 
Frame rate & up to 500 fps & 100 fps \\ \hline 
Pixel clock & up to 2.5 MHz & 110 kHz  \\ \hline 
Read noise & $<$ 10 e- &  \\ \hline 
Linearity & $<$ 1\% after flat-fielding &  \\ \hline 
Uniformity & $<$ 10\% &  \\ \hline 
\end{tabular}
\vspace{0.2cm}
\caption{\label{tab:detreq} Infrared wavefront sensor detector requirements.}
\end{center}
\end{table}

After an intense market survey, and contact with different manufacturers, the following options were investigated in detail:  the HAWAII 1 and the new HAWAII 1-RG and 2-RG devices from Teledyne. The HAWAII-1 focal plane array was discarded at an early stage of the Phase A study because the desired frame rates (up to 500 fps) would not have been achieved (the maximum frame rates obtained would be 25 Hz for the largest frame, and 115 Hz for the optimized frame).

\begin{table}[ht]
\begin{center}
\begin{tabular}{|p{1.5in}|p{1.0in}|p{1.2in}|p{1.0in}|} \hline 
Detector Type & Frame size (pix) & Pixel clock (kHz) & Frame rate (Hz) \\ \hline 
HAWAII-1 & 4x 100x100 & 250 & 25 \\ \hline 
 & 4x 60x6x6 & 250 & 115 \\ \hline 
HAWAII-1RG / 2RG & 4x 100x100 & 2500 & 500 \\ \hline 
 & 4x 100x100 & 1000 & 200 \\ \hline 
 & 4x 100x100 & 500 & 100 \\ \hline 
 & 4x 60x6x6 & 1000 & 926 \\ \hline 
 & 4x 60x6x6 & 550 & 509 \\ \hline 
 & 4x 60x6x6 & 250 & 230 \\ \hline 
 & 4x 60x6x6 & 110 & 102 \\ \hline 
\end{tabular}
\vspace{0.2cm}
\caption{Comparison HAWAII-1 vs. HAWAII-1RG/2RG.}
\end{center}
\end{table}  

On the other hand, the new -RG devices have better quantum efficiencies\cite{Finger2004a} and allow clocking frequencies up to 5MHz\cite{Loose2007a}, what fulfils the frame rate requirements.  Since their video output signals are organized into stripe-like channels that can be read out in parallel, and the clocking direction can also be set, an area selected between 2 channels can be read as twice as fast.  In this way, each of the 4 beams can be spread over 2 channels and the readout time can thus be reduced by a factor two. In order to minimize line skipping, the beams should be positioned either close to the top or bottom edge of the array.

From the operating point of view of the GRAVITY WFS, the performance of the HAWAII-1RG and the HAWAII-2RG arrays is the same, and therefore both could be used. There are at least 8 Channels needed to accommodate the 4 beams and both detectors fulfil this requirement.  Since the HAWAII-2RG is already supported by ESO, the baseline design is based on it.  

   \begin{figure}[ht]
   \begin{center}
   \begin{tabular}{c}
   \includegraphics[width=10cm]{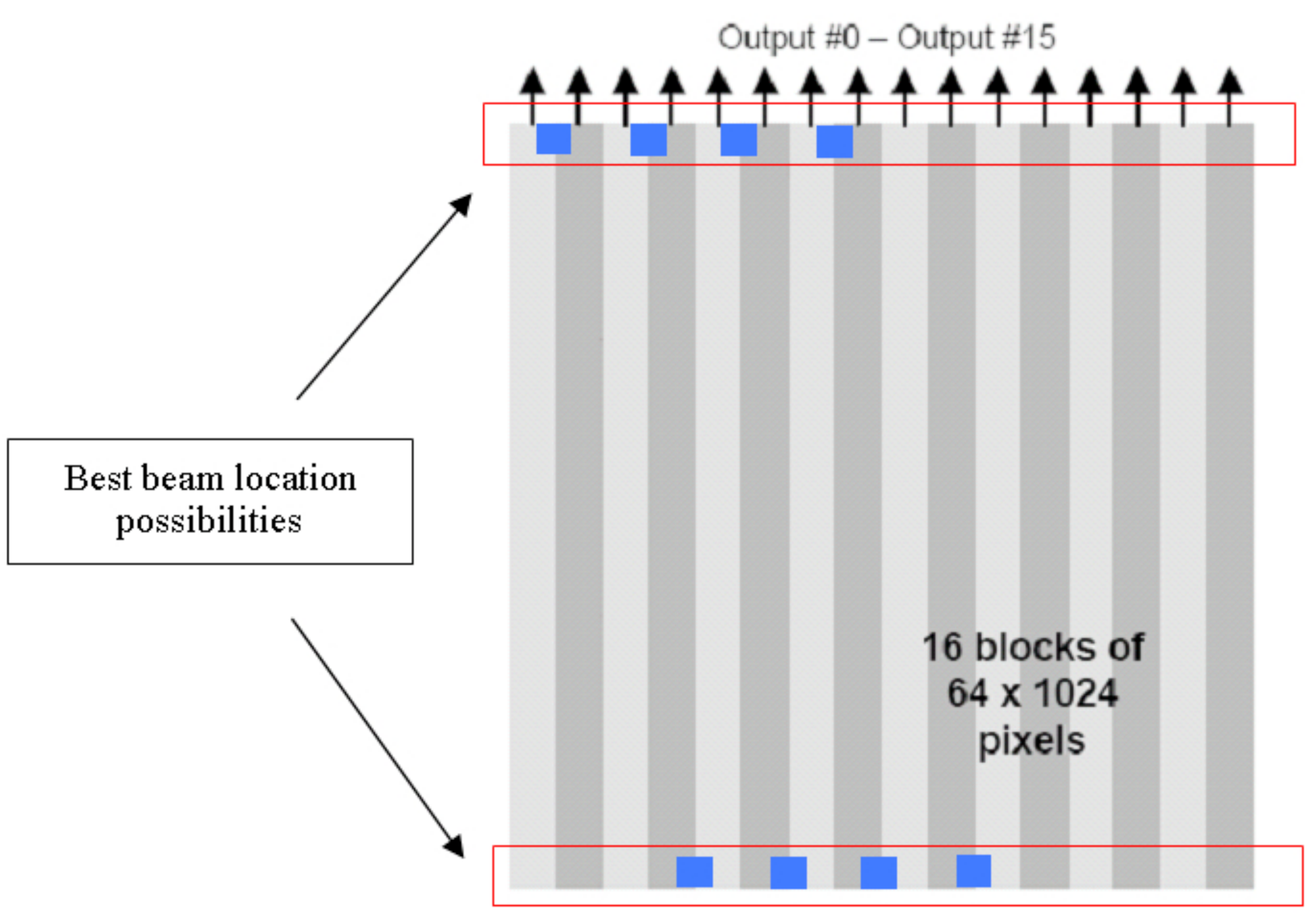}
   \end{tabular}
   \end{center}
   \caption[Schematic] 
   { \label{fig:BestDetLocation} 
HAWAII-1RG channel distribution with best beam location possibilities.}
   \end{figure} 

\subsection{Deformable mirrors}
To complete a fully working adaptive optics system, our baseline design includes the usage of 
existing VLTI infrastructure, in particular the MACAO deformable mirrors. MACAO uses 
bimorph deformable mirrors with 60 actuators.

\section{Phase A adaptive optics performance simulations}
\subsection{Estimation of the optimal bandpass}
The choice of the bandpass has to take into account: the background spectrum, the object spectrum, the read-out noise of the detector (RON) ,the sampling frequency $F_S$. The bandpass is optimized by choosing $(\lambda_1,\lambda_2)$ that make the SNR maximum:  $\int^{\lambda_2}_{\lambda_1} O(\lambda)d\lambda/\sqrt{\int^{\lambda_2}_{\lambda_1} B(\lambda)d\lambda + F_S.RON^2}$, where $O(\lambda)$ is the object spectrum per second, $B(\lambda)$ the background spectrum per second, $F_S$ the sampling frequency and RON the detector read-out noise.

In the Galactic Centre case, modelling IRS7 (see fig.~\ref{fig:GC_AO}) by blackbody with T=840~K, we have studied the optimal bandpass versus sampling frequency (100 to 200 Hz) and RON (from 5 to 20 e$^-$) for different VLTI temperatures. We have assumed a VLTI  transmission of 0.1 (ie a VLTI emission of 0.9). As a conclusion,  (i) the J band is not absolutely required, (ii) a good compromise could be to use the band 1.44-2.45 $\mu$m, (iii) a blocking filter in the 1.80-1.96 $\mu$m region, where the atmosphere becomes opaque , is not required.

\subsection{Dimensioning the wavefront sensor}
We have considered IRS7 as the AO reference source and the optimal bandpass defined above. Building on experience with NAOS, we have chosen a pixel size equal to the width of the diffraction pattern of a subaperture, i.e. ($\lambda/d_{subap}$). This is also equal to the number of subapertures times the telescope diffraction limit. We have aslo assumed the following parameters: $T_{VLTI}$=0.1, $T_{WFS}$=0.85, $T_{filter}$=0.9, QE=0.6, $t_{VLTI}$=12$^\circ$C  (corresponding to a background=26500 e$^-$/m$^2$/s/arcsec$^2$), wind speed=12 m/s corresponding to $\tau$=3ms in the visible. We have then computed the error budget (fitting, aliasing, temporal, noise, anisoplanatism errors) for different reference source magnitudes, different seeing conditions, different RON and $F_S$, and different number of Shack Hartmann subpupils. Results are given in the following figures, expressed in rd$^2$. The Strehl ratio is computed on the PSF image: it has been calculated by a Monte-Carlo code, simulating phase screens, compensated by the influence functions of the MACAO mirror, introducing noise (propagated though the system control matrix taking into account the Hartmann geometry), aliasing properties, servo-lag and anisoplanatism.

In these simulations, the sampling frequency and loop bandwidth have been optimized in order to 
minimize the phase error. In particular, at low flux, the bandwidth has been reduced in order to 
smooth the noise: this results in a higher temporal error, although the sum of (noise + temporal) 
is minimized. 

The SR estimates differ slightly from $\exp(-\sigma^2)$ at low values because the expression $SR=\exp(-\sigma^2)$ is just an approximation and above all because the various error sources (fitting, aliasing, temporal, noise, anisoplanatism) are correlated between each other, and thus do not simply add up together.

   \begin{table}[ht]
   \begin{center}
   \begin{tabular}{c}
   \includegraphics[width=12cm]{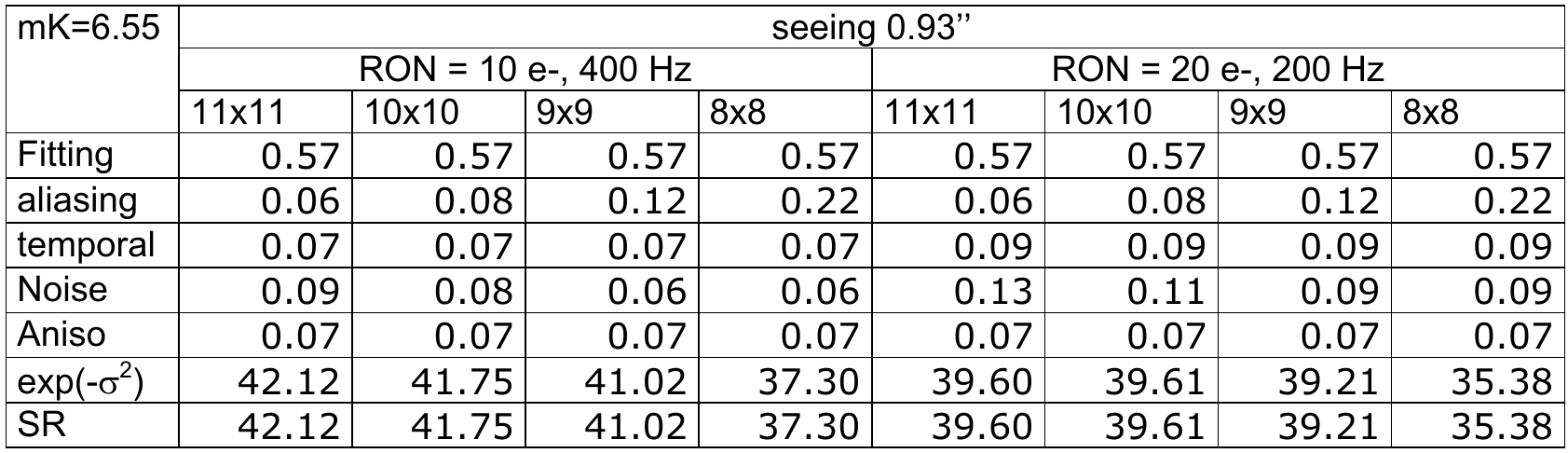}
   \end{tabular}
   \end{center}
   \caption[Error budget for IRS7 in median seeing conditions] 
   { \label{tab:eb1} 
Error budget for IRS7 in median seeing conditions.}
   \end{table} 
   \vspace*{-1.0cm}
   \begin{table}[ht]
   \begin{center}
   \begin{tabular}{c}
   \includegraphics[width=12cm]{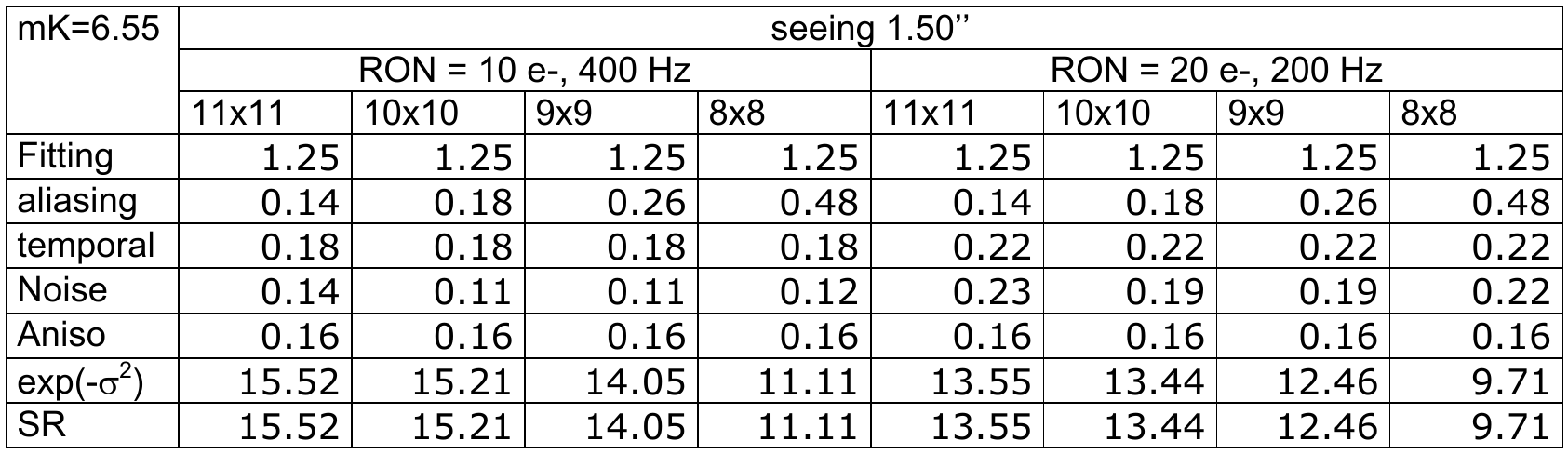}
   \end{tabular}
   \end{center}
   \caption[Error budget for IRS7 in bad seeing conditions] 
   { \label{tab:eb2} 
Error budget for IRS7 in bad seeing conditions.}
   \end{table} 
   \vspace*{-1.0cm}
   \begin{table}[ht]
   \begin{center}
   \begin{tabular}{c}
   \includegraphics[width=12cm]{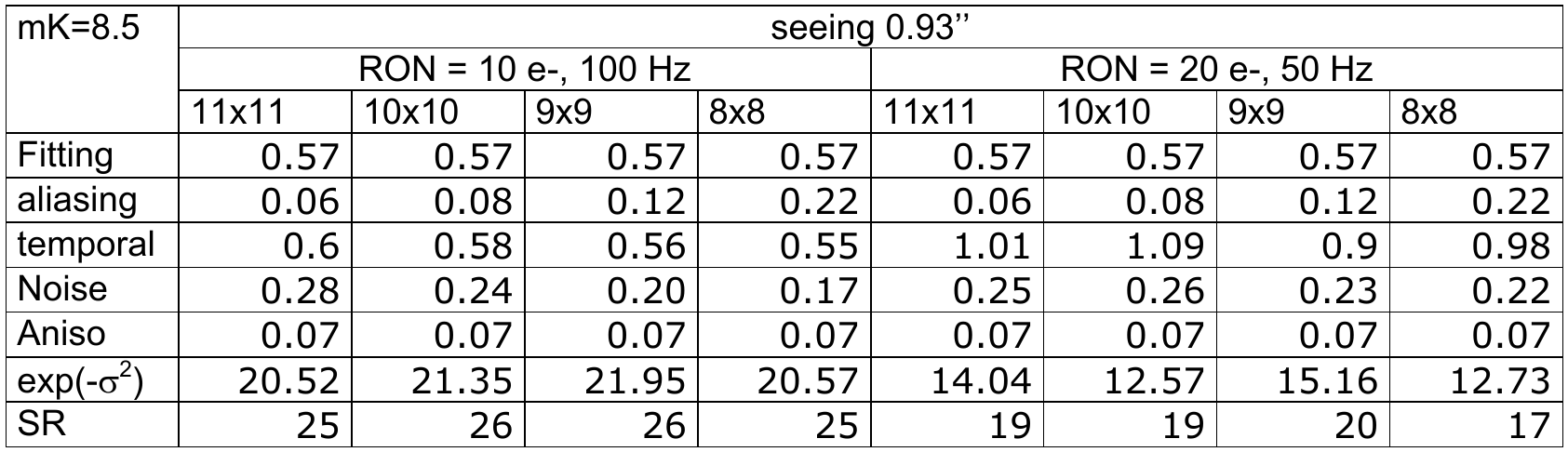}
   \end{tabular}
   \end{center}
   \caption[Error budget for mK=8.5 in median seeing conditions] 
   { \label{tab:eb3} 
Error budget for mK=8.5 in median seeing conditions.}
   \end{table} 
   \vspace*{-1.0cm}
   \begin{table}[ht]
   \begin{center}
   \begin{tabular}{c}
   \includegraphics[width=12cm]{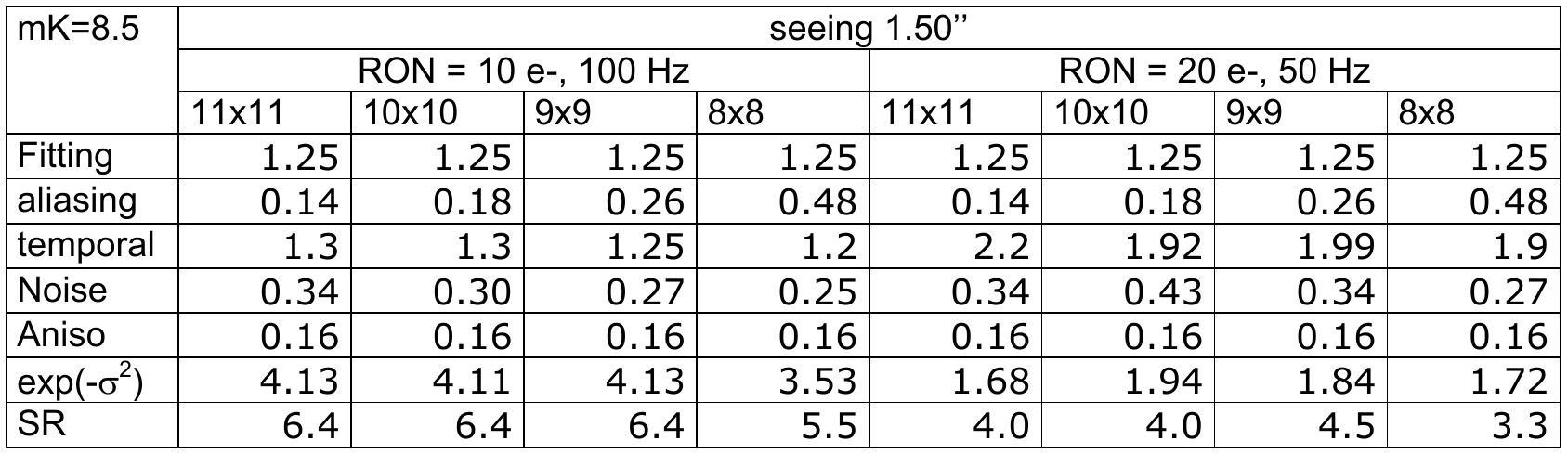}
   \end{tabular}
   \end{center}
   \caption[Error budget for mK=8.5 in bad seeing conditions] 
   { \label{tab:eb4} 
Error budget for mK=8.5 in bad seeing conditions.}
   \end{table} 
   \vspace*{-1.0cm}
   \begin{table}[ht]
   \begin{center}
   \begin{tabular}{c}
   \includegraphics[width=6.0cm]{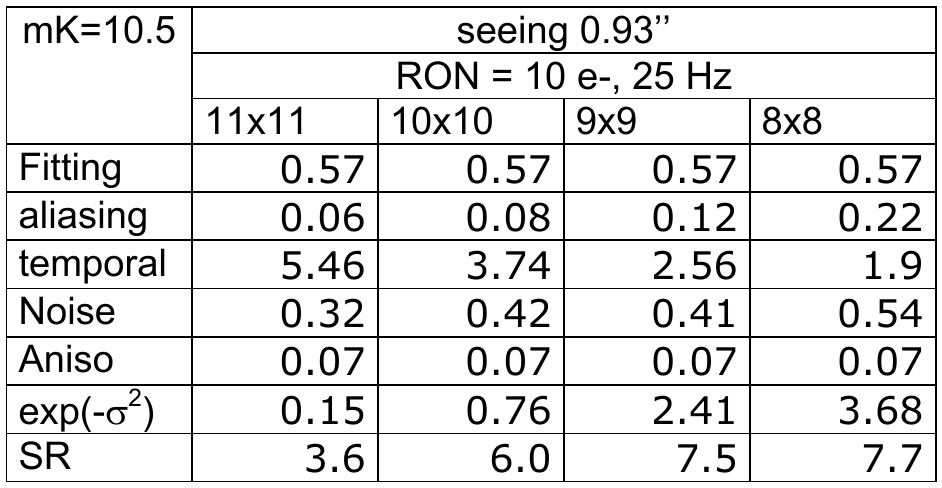}
   \end{tabular}
   \end{center}
   \caption[Error budget for mK=10.5 in median seeing conditions] 
   { \label{tab:eb5} 
Error budget for mK=10.5 in median seeing conditions.}
   \end{table} 
 \clearpage

From these simulations, we concluded that the top-level requirements are met for the mK=7 cases and closely met for the mK=10 case. This holds for the above mentioned configuration of 10 e- RON, sampling frequency and number of sub-apertures depending on brightness of the guide star. For the GC case the top level requirements are always fulfilled, even with a RON of 20 e-.
\section{Optical design}
The optical design of the WFS follows the general concept of providing one single system for all four beams, using UT1-4 or AT1-4. These beams when leaving the switch yard are separated by 240mm, each pupil diameter is 18mm. 

\begin{figure}[ht]
   \begin{center}
   \begin{tabular}{c}
   \includegraphics[width=16cm]{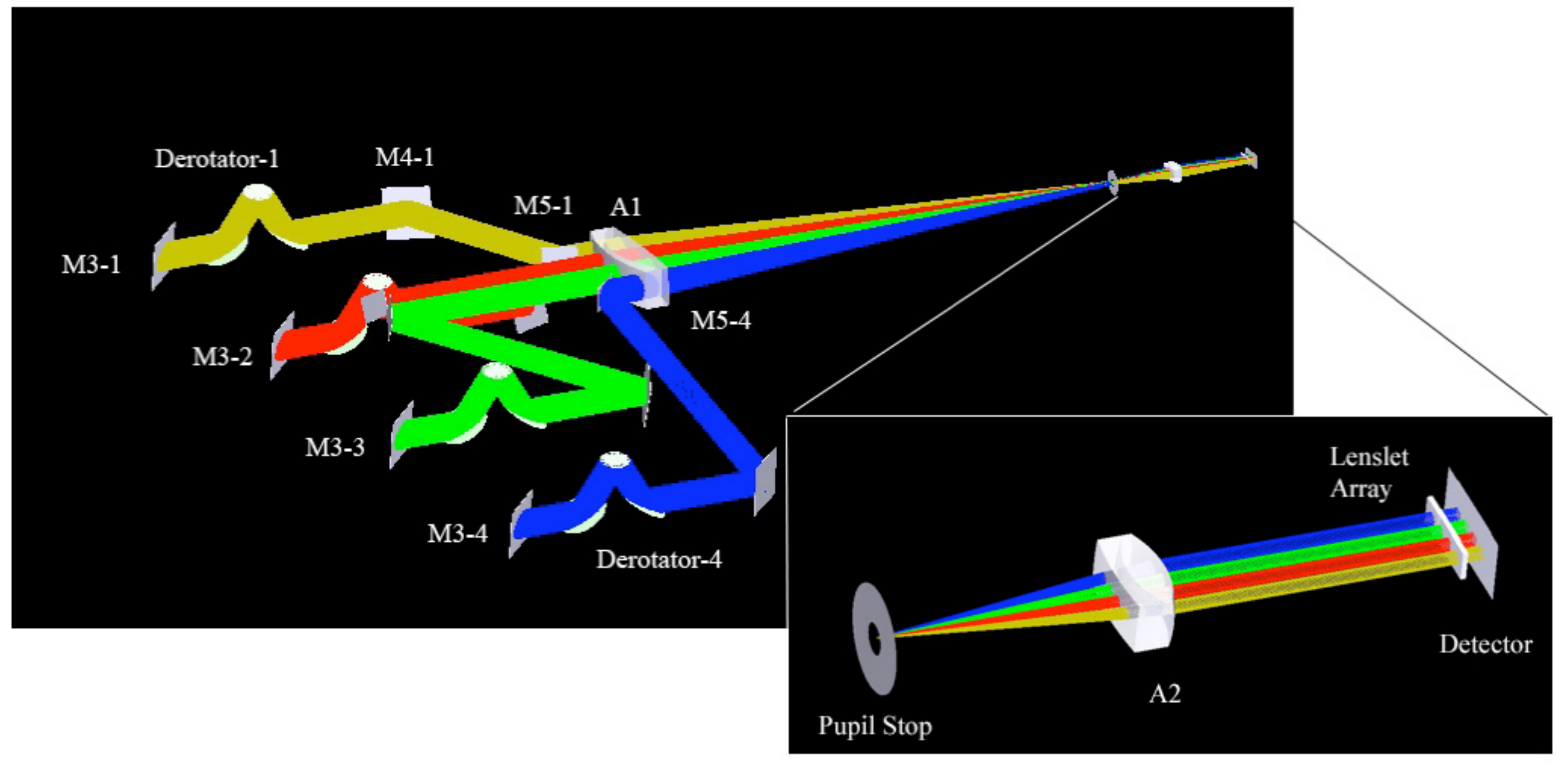}
   \end{tabular}
   \end{center}
   \caption[Schematic] 
   { \label{fig:OpticalDesign_Schematic} 
Optical layout of the WFS: the right sub-image shows details near the lenslet array.}
   \end{figure}

The first component of the WFS optics (see Fig.~\ref{fig:OpticalDesign_Schematic}) is the pupil-de-rotator, one individual component for all four telescopes. This device is realized by a carefully aligned K-mirror assembly, the individual mirrors are protected silver coated. 

The following flat mirrors M4-1 to M4-4 together with M5-1 to M5-4 are feeding the four beams into the common optics. At the same time, M4-1 to M4-4 serve as field selection unit, the intermediate pupil image position has to be adjusted to these mirrors. 

The following Achromat A1 is producing a common focused image centered to a cold field stop. This field stop reduces stray-light and ghost images down to a negligible value. A small achromat A2 re-images the pupil onto the lenslet array, which is mounted directly in front of the detector array. 
The current optical design of these two achromatic lenses is producing a pupil diameter of 2.4mm, which has to be adjusted to the finally chosen pixel size, number of sub-pupils and detector array channels.  

\newpage
\section{Instrument control electronics}
\subsection{Requirements}
The  instrument control electronics (ICE) realizes all control and monitoring tasks for the GRAVITY instrument. It controls several electro-mechanisms, sensors, cryostats and electronics cabinets. GRAVITY uses a large number of electro-mechanical devices, which increases the complexity of the control electronics. In order to keep the electronics design simple and allow for easy maintenance, standard actuators, sensors and control devices from same model are used. Whenever possible, ESOÕs standard components are used. 
\subsection{Architecture}
The ICE is divided into several subsystems, reflecting the functional subunits of GRAVITY. The control electronics architecture is based on the computer system architecture of ESO. The instrument workstation is connected to the control electronics via LAN sub-net. Inside the cabinet the CPU of a Local Control Unit (LCU) communicates directly with other devices via VME-bus. Non VME devices communicate with the CPU board via RS232 interface. 

Three LCUs are present in the system. One LCU controls the AO system and the Switch-yard, the other two LCUs control the Fibre Coupler including the Field Selector, the Beam Combiner, the Spectrometers, and the Metrology system. The figure below shows the control system architecture for the AO system.
   \begin{figure}[ht]
   \begin{center}
   \begin{tabular}{c}
   \includegraphics[width=12cm]{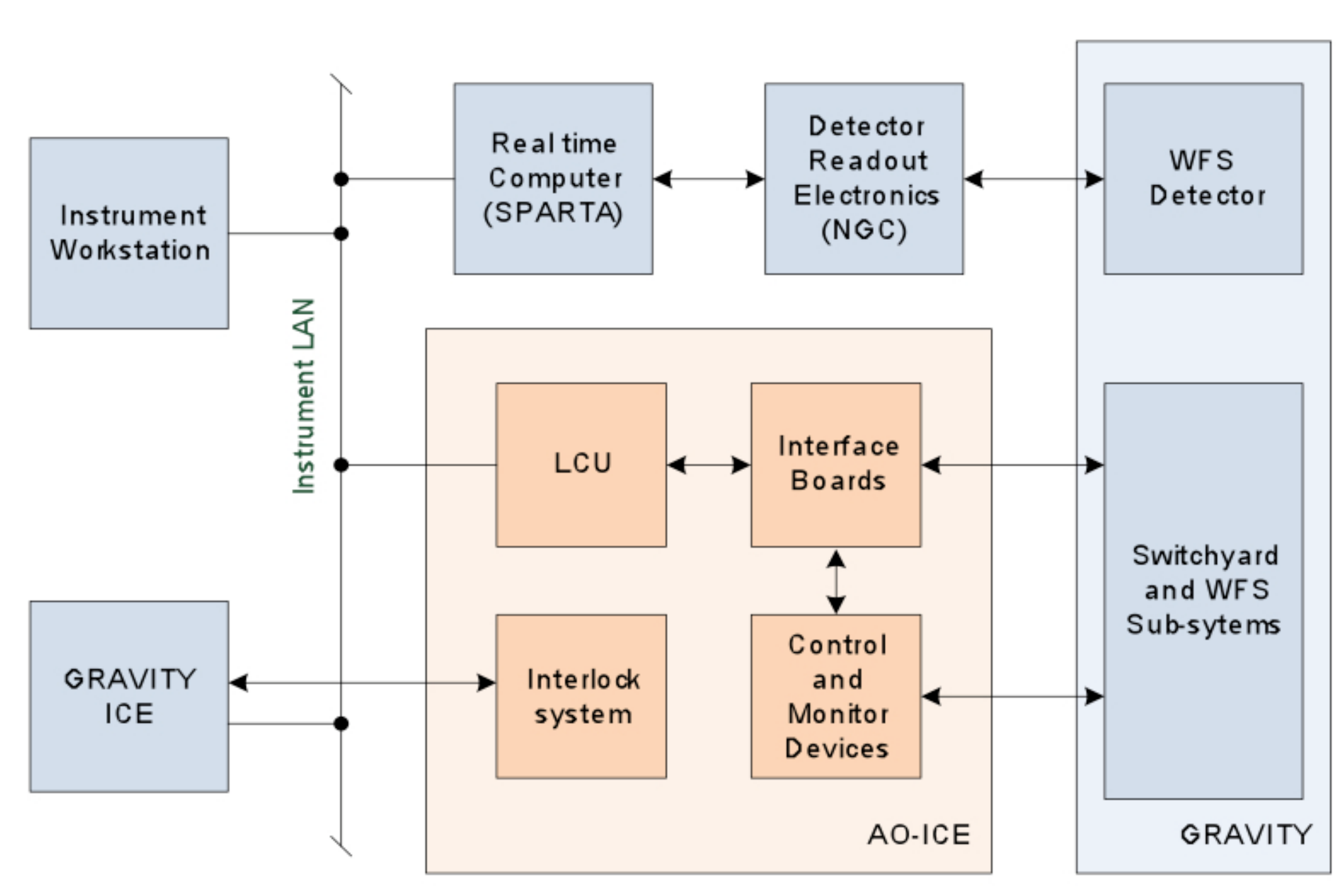}
   \end{tabular}
   \end{center}
   \caption[Schematic] 
   { \label{fig:ICE_Schematic} 
Control electronics architecture of GRAVITY.}
   \end{figure} 


\section{VLTI internal seeing and its impact on the location of the wavefront sensor}
\label{sec:tunnel}
In order to characterise the wavefront aberrations caused by turbulence in the VLTI tunnels and light ducts, a measurements campaign was started in the last week of August 2007. The setup consists of an commercial Shack-Hartmann (SH) sensor (ÒWavescopeÓ, manufactured by Adaptive Optics Associates) with control PC and an additional frame grabber PC capable of recording the spot pattern of the SH sensor at a frame rate of 30\,Hz. The control PC of the SH sensor is used for alignment of the Wavescope unit, checking of illumination level and integration times, and to decompose the recorded wavefront measurements into Zernike modes. While this system allows data acquisition only at a maximum speed of 15\,Hz, the external frame grabber solution directly records the video signal of the WavescopeÕs internal camera at a full frame rate of 30\,Hz. The recorded spot patterns can be fed back to the analysis software of the Wavescope unit to allow the same data reduction at doubled temporal sampling. For the example measurement discussed below, the HeNe laser beacon at the Coud\'{e} focal station of the VLT  was used as light source. The used delay line number 2 was set-up to compensate an optical path length of 100\,m. In order to match the beam size (80\,mm) to the entrance pupil size of the SH sensor (10\,mm), a single converging lens was used. The SH sensor was equipped with a microlens array with 480\,$\mu$m pitch size, resulting in useful data in $\sim$250-300 sub-apertures.
 
\subsection{ Spatial wavefront characteristics}

\noindent In case of Kolmogorov turbulence, the variance of the wavefront $\sigma_{\Phi}^2$ is related to the aperture size D, the Fried parameter r$_0$, and the sum of the Zernike coefficient $a_i$ variances by:

\begin{center}
\begin{math}
  \sigma_{\Phi}^2 = 1.0299 \left( \frac{D}{r_0}\right) ^{5/3} = \sum < (a_i - \overline{a}_i)^2 >
\end{math}
\end{center}

\noindent This assumes an infinite outer scale L$_0$, and it has to be decided from case to case if this assumption is valid for the given conditions. If the outer scale is finite but still well larger than the Fried parameter r$_0$, the above formula can still be applied by using only the higher order Zernike modes, in the simplest case starting with the focus term. In this case the factor 1.0299 reduces to 0.134, reflecting that the major part of wavefront variance is contained in the tip/tilt modes. By comparing the two values for r0 derived in this way, one can already test if Kolmogorov statistics with infinite outer scale or the 
van-K\'{a}rm\'{a}n model with finite L$_0$ allows the best representation of the measured turbulence. 

\noindent The following figure shows the temporal evolution of the wavefront rms with and without tip-tilt aberrations during a 5000\,s measurement with 30\,Hz sampling rate. Static aberrations have been removed.

   \begin{figure}[ht]
   \begin{center}
   \begin{tabular}{c}
   \includegraphics[width=8cm]{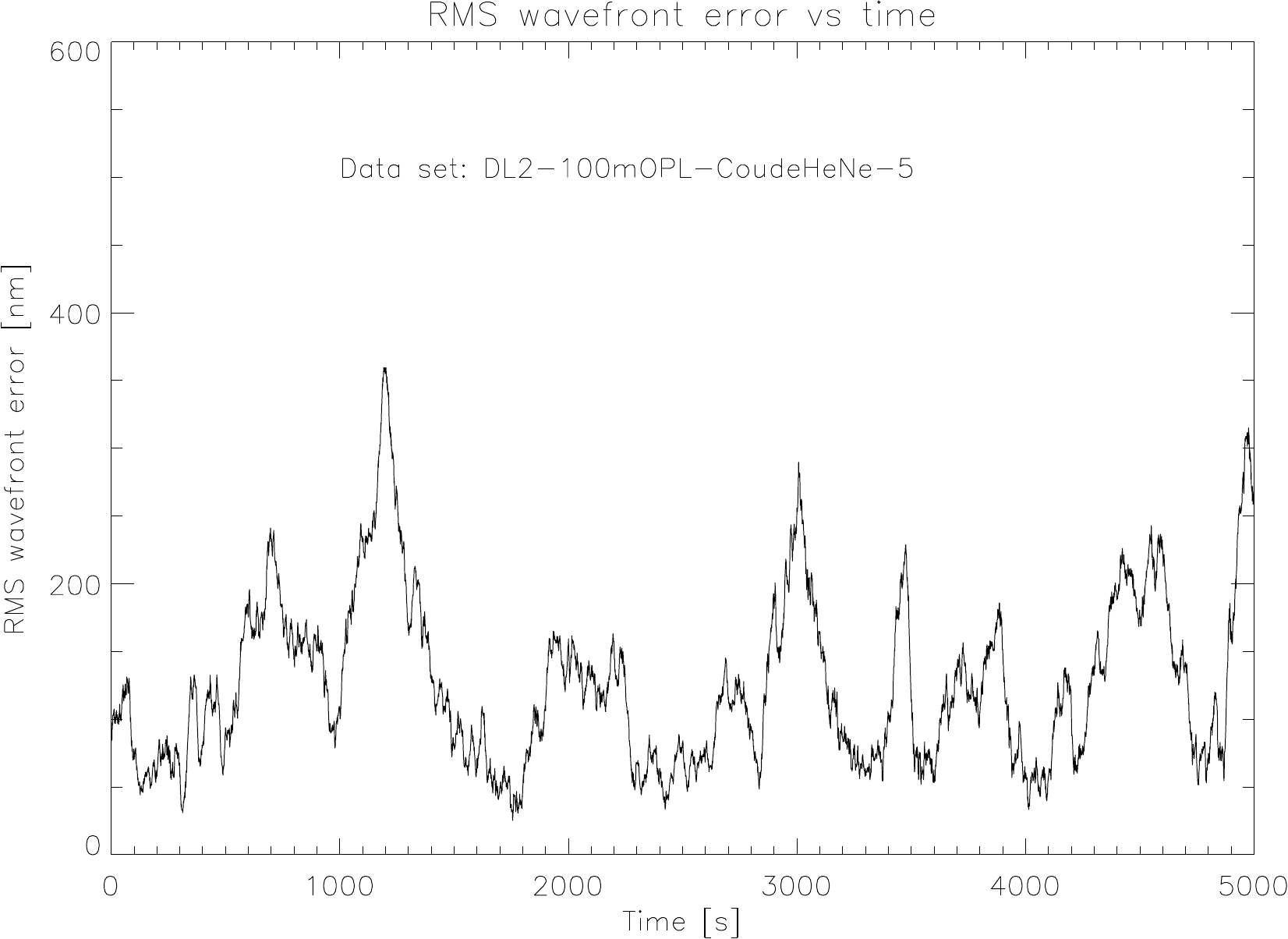}
    \includegraphics[width=8cm]{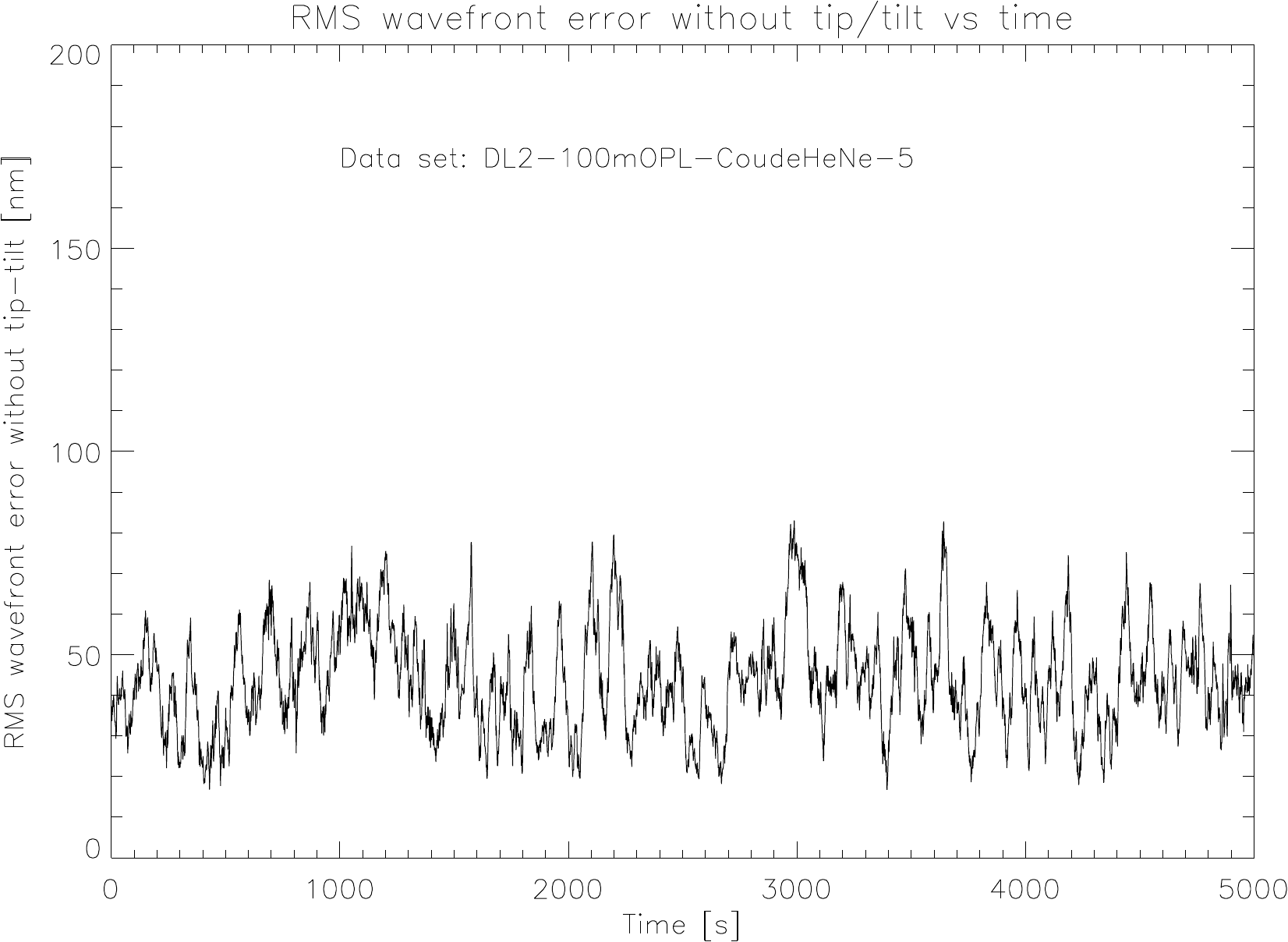}
   \end{tabular}
   \end{center}
   \caption[WFSrms] 
   { \label{fig:WFS_RMS} 
Wavefront rms with static aberrations excluded (left) and without tip-tilt (right).}
   \end{figure} 

\noindent The root mean square (rms) value of the wavefront $\sigma_{\Phi} $ is 126\,nm, equivalent to 1.25\,rad at the laser's wavelength or 0.36\,rad  in \textit{K}-band. The resulting Strehl ratio -- assuming an otherwise perfect optical system -- can be approximated by $SR = \exp^{-\sigma_{\Phi}^2}$.

\noindent The numbers for this example are SR=0.21 and SR=0.88 in the red and in \textit{K}-band, respectively. At the laser wavelength of 632.8\,nm, the corresponding Fried parameter r$_0$ for this series is 6.3\,cm. Taking the beam compression ratio of $\approx$100 into account, this translates to an effective on-sky r$_0$ of 6.3\,m. At 2.2$\mu $m, the corresponding value would be 28.1\,m (using the relation r$_0 \sim \lambda^{6/5}$), well larger than the telescope diameter. 

\noindent If a perfect tip/tilt-correction can be accomplished, the remaining wavefront rms can be evaluated by considering only the higher Zernike modes, starting with the focus term (see right plot of Fig.~\ref{fig:WFS_RMS}). In this case, the mean wavefront rms reduces to 43.9\,nm, giving an effective on-sky value for r$_0$ of 29\,m in the red and 130\,m in K-band.

\noindent The fact that the wavefront rms after removal of tip and tilt was reduced by a factor of 2.87 shows that the turbulence is quite similar to
Kolmogorov turbulence where a factor of 2.8 is expected (variance reduces from 1.0299 to 0.134).

\subsection{Temporal wavefront characteristics}

\begin{figure}[ht]
   \begin{center}
   \begin{tabular}{c}
   \includegraphics[width=8cm]{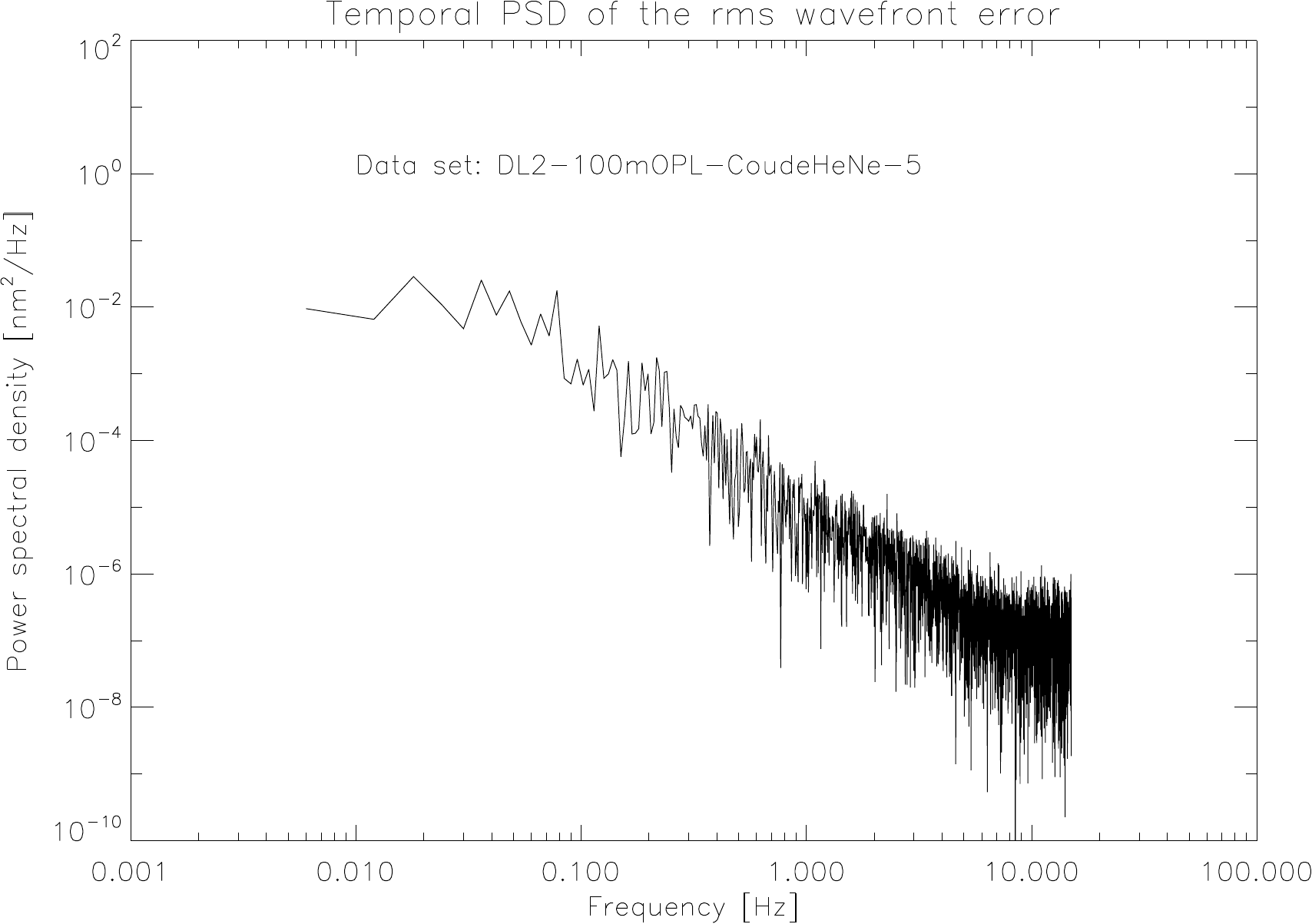}
    \includegraphics[width=8cm]{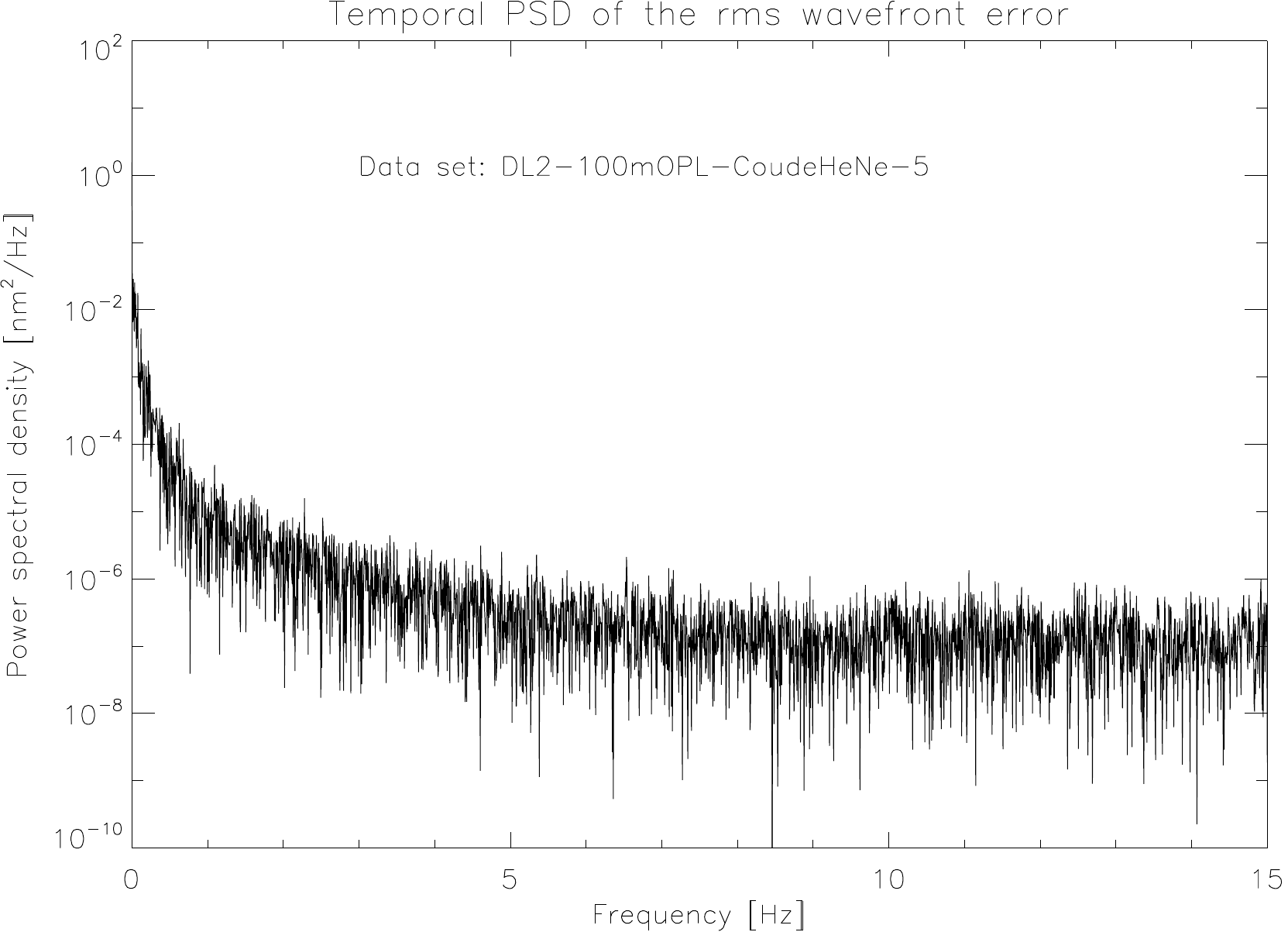}
   \end{tabular}
   \end{center}
   \caption[WFSpsd] 
   { \label{fig:wavefront_PSD} 
Temporal power spectrum of the wavefront RMS. The left plot uses logarithmic scaling on both axes, the right one only for the power.}
   \end{figure}

\noindent Figure~\ref{fig:wavefront_PSD} shows the temporal power spectrum of the total wavefront rms evolution. 

\noindent

\section{Summary}
Adding near-infrared wavefront sensing capabilities to the VLTI will increase the number of possible science targets
 for VLTI instruments using adaptive optics. While compensating high-order atmospheric turbulence up to the Coud\'{e} focal station of the UTs (location of the deformable mirror) there are currently two options available to compensate residual low-order aberrations up to the VLTI laboratory with the IRIS wavefront sensor. One uses the XY table of the MACAO wavefront sensor\cite{Schoeller2007} as an indirectly and rather slow tilt compensating device, the other option uses fast piezo driven steering mirrors inside the VLTI laboratory instrument feeding optics (currently only available for AMBER and PRIMA). For GRAVITY we investigated the power of high order aberrations introduced between the Coud\'{e} deformable mirror (M8) and the interferometric lab. With typically around 50\,nm rms it is questionable whether they need to be compensated. For the residual low-order aberrations, a compensation as close as possible to the science instrument seems mandatory in order to stabilize the beams within fibers and other position sensitive devices.

\acknowledgments     
The authors like to thank Philippe Gitton and Pierre Haguenauer, who measured and supported the measurements of the VLTI tunnel seeing in the
context of the GRAVITY phase A study. 


\bibliography{report}   
\bibliographystyle{spiebib}   

\end{document}